\documentclass[twocolumn,secnumarabic,amssymb, amsmath, nofootinbib,tightenlines,
nobibnotes, aps, prl,epsfig]{revtex4}
\usepackage{graphicx}% Include figure files
\usepackage{dcolumn}% Align table columns on decimal point
\usepackage{bm}% bold math
\begin{document}
\preprint{APS/123-QED}
\title{Ratio of the structure functions and the color dipole model bound}% Force line breaks with \\

\author{G.R.Boroun}%
 \email{grboroun@gmail.com; boroun@razi.ac.ir }
 \author{B.Rezaei}%
 \email{ brezaei@razi.ac.ir }
\affiliation{ Physics Department, Razi University, Kermanshah
67149, Iran}% \textbackslash\textbackslash
\date{\today}% It is always \today, today,
\begin{abstract}
%%%%%%%%%%%%%%%%%%%%%%%%%%%%%%%%%%%%%%%%%%%%%%%%%%%%%%%
We observe that the DGLAP evolution equations at NNLO analysis
predicts a ratio of the structure functions in region of small
Bjorken variable $x$. The ratio $F_{L}(x,Q^{2})/F_{2}(x,Q^{2})$ is
obtained and compared with the prediction of the dipole model and
HERA data. In particular we show that this ratio is lower than
dipole model bound at high-$Q^{2}$ values and  it is higher at
low-$Q^{2}$ values . Then the effect of adding a higher twist term
to the description of the ratio $F_{L}(x,Q^{2})/F_{2}(x,Q^{2})$
 for $Q^{2}< 20~GeV^{2}$ is investigated. Also the bounds are discussed
by including charm distribution on $F_{L}/F_{2}$. We discuss,
furthermore, how this ratio can be determine the proton structure
function with respect to the
reduced cross section at high-$y$ values.\\

%%%%%%%%%%%%%%%%%%%%%%%%%%%%%%%%%%%%%%%%%%%%%%%%%%%%%%%
\end{abstract}
 \pacs{***}%PACS, the Physics and Astronomy
                              %Classification Scheme.
 %Use showkeys class option if keyword
\maketitle
\tableofcontents
%**********************************************************
\subsection{1. Introduction}
Measurements of the inclusive deep inelastic scattering (DIS)
cross section have been pivotal in the development of the
understanding of strong interaction dynamics [1-5]. The cross
section in this measurement depends on two structure function
$F_{2}$ and $F_{L}$. Indeed these functions are depend on  the
kinematic variables $x$ and $Q^{2}$. The structure functions
obtained from these experiments have helped develop the
description of hadrons. Hadrons are composite objects from the
quarks and gluons at low and high-$x$ values. The longitudinal
structure function $F_{L}(x,Q^{2})$ comes as
$F_{L}(x,Q^{2})=F_{2}(x,Q^{2})-2xF_{1}(x,Q^{2})$, where
$F_{2}(x,Q^{2})$ is the transverse structure function and it can
be expressed as a sum of the quark-antiquark momentum
distributions $xq_{i}(x)$ weighted with the square of the quark
electric charges $e_{i}$:
$F_{2}=x\sum_{i}e_{i}^{2}(q+\overline{q})$. Also $F_{L}$ is
directly dependent on the gluon distribution
 and it is proportional to the running coupling constant $\alpha_{s}$.\\
In the one-photon exchange approximation the neutral current
 reduced cross section is defined as
\begin{eqnarray}
\sigma_{r}(x,Q^{2})=F_{2}(x,Q^{2})[1-\frac{y^{2}}{Y_{+}}\frac{F_{L}(x,Q^{2})}{F_{2}(x,Q^{2})}],
\end{eqnarray}
where $Y_{+}=1+(1-y)^2$, $y={Q^{2}}/{xs}$ is the inelasticity and
$s$ is the center-of-mass  squared energy of  incoming electrons
and protons respectively.  The transverse and longitudinal
structure functions, $F_{2}(x,Q^{2})$ and $F_{L}(x,Q^{2})$ , are
related to the transverse and longitudinal virtual photon
absorption cross section, $\sigma_{T}$ and $\sigma_{L}$. It is
convenient to define the structure functions as follows
\begin{eqnarray}
F_{2}(x,Q^{2})&=&\frac{Q^{2}}{4\pi^{2}\alpha_{em}}(1-x)[\sigma_{T}(x,Q^{2})+\sigma_{L}(x,Q^{2})],\nonumber\\
F_{L}(x,Q^{2})&=&\frac{Q^{2}}{4\pi^{2}\alpha_{em}}(1-x)\sigma_{L}(x,Q^{2}).
\end{eqnarray}
Where the contribution of $F_{L}$ to  reduced cross section (
Eq.(1)) is significant only at high value of the inelasticity $y$,
in spite of the fact that data on $F_{L}$ are generally difficult
to extract from the cross section measurements.\\
In the first approximation of the parton model, the longitudinal
structure function  is equal identically zero but in actual DIS
experiments should be nonzero since it arises from gluon
corrections. Therefore  $ {F_{L}(x,Q^{2})}$  behavior is
dependence on values of $Q^{2}$. This behavior in the dipole
picture [6] for DIS $F_{L}$  is nonzero. In the dipole model a
strict bound for the ratio of
$\frac{F_{L}(x,Q^{2})}{F_{2}(x,Q^{2})}$ is defined as [7-8]
\begin{eqnarray}
\frac{F_{L}(x,Q^{2})}{F_{2}(x,Q^{2})}\leq 0.27.
\end{eqnarray}
Based on the dipole formulation of the $\gamma^{*}p$ scattering
[9], the standard formulae for $F_{2}$ and $F_{L}$ are defined by
\begin{eqnarray}
F_{2}(x,Q^{2})&=&\frac{Q^{2}}{4\pi^{2}\alpha_{em}}(1-x)\sum_{q}\int d^{2}r [w^{(q)}_{T}(r,Q^{2})\nonumber\\
&&+w^{(q)}_{L}(r,Q^{2})]\widehat{\sigma}^{(q)}(r,\xi),\\
F_{L}(x,Q^{2})&=&\frac{Q^{2}}{4\pi^{2}\alpha_{em}}(1-x)\sum_{q}\int
d^{2}r w^{(q)}_{L}(r,Q^{2})
\widehat{\sigma}^{(q)}(r,\xi),\nonumber
\end{eqnarray}
where $w^{(q)}_{T,L}$ are the probability densities for the
virtual photon splitting into a $q\overline{q}$ pair and
$\widehat{\sigma}$ is the dipole cross section which describes the
interaction of the dipole with the proton. This cross section
depends on $r$ where it is the
transverse separation of the quarks in the quark-antiquark pair, and $\xi $ is an energy variable in this formalism.\\
The bound for the ratio $\frac{F_{L}(x,Q^{2})}{F_{2}(x,Q^{2})}$
defined [10-11]
\begin{eqnarray}
g(Q,r,m_{q})=\frac{w^{(q)}_{L}(r,Q^{2})}{w^{(q)}_{T}(r,Q^{2})+w^{(q)}_{L}(r,Q^{2})},
\end{eqnarray}
where $m_{q}$ is the mass of the quark $q$. It was shown in
literatures that for all $Q\geq0$, $r\geq0$ and $m_{q}\geq0$ the
bound (5) for the ratio ${F_{L}}/{F_{2}}$ is valid.\\
The paper is organized as follows. In section 2 we describe a
formalism for the solution of DGLAP evolution equations [12] at
NNLO analysis. We suggest an evolution method for the ratio
$\frac{G(x,Q^{2})}{F_{2}^{s}(x,Q^{2})}$ in this section. Then the
ratio $\frac{F_{L}}{F_{2}}$ from the Altarelli-Martinelli equation
[13] would be obtained and compared with HERA data and with the
color dipole model bound. The results and discussions of our
predictions presented in section 3. A connection between the
structure function from the DGLAP evolution equations with the
color dipole model (CDM) discussed in this section. Then allows
one to draw conclusions about the role of higher twist effects in
the ratio of structure functions. An influence of heavy quark
contribution to the ratio ${F_{L}}/{F_{2}}$ is discussed in
section 4. We conclude in
section 5.\\
%%%%%%%%%%%%%%%%%%%%%%%%%%%%%%%%%%%%%%%%%%%%%%%%%%%%%%%%%%%
\subsection{2. Formalism}
\subsection{2.1. The ratio $\frac{G(x,Q^{2})}{F_{2}^{s}(x,Q^{2})}$}
The DGLAP evolution equations for the singlet and  gluon density
in the standard form are given by
\begin{eqnarray}
\frac{d}{d lnQ^{2}} \left[\begin{array}{c}
 q_{s}(x,Q^{2}) \\
 g(x,Q^{2})
 \end{array}
 \right]=\left[
 \begin{array}{cc}
 P_{qq} & P_{qg}\\
 P_{gq} & P_{gg}
 \end{array}
 \right]{\otimes}
\left[\begin{array}{c}
 q_{s}(x,Q^{2}) \\
 g(x,Q^{2})
 \end{array}
 \right]
\end{eqnarray}
which emphasized that quark and gluon densities are coupled. The
convolution express the possibility that a parton $i$ with
momentum fraction $x$ may originate from the branching of a parent
parton $j$ of the higher momentum fraction $y$ ($P_{ij}$ is the
splitting function) and is defined by $P_{ij}{\otimes}
\emph{f}_{j}=\int_{x}^{1}\frac{dy}{y}P_{ij}(\frac{x}{y})\emph{f}_{j}(y,Q^{2})$.
The singlet quark density of a hadron is given by
\begin{eqnarray}
q_{s}(x,Q^{2})=\sum_{i=1}^{N_{f}}[q_{i}(x,Q^{2})+\overline{q}_{i}(x,Q^{2})],
\end{eqnarray}
where $q_{i}$ and $\overline{q}_{i}$ represent the number
distribution of quarks and antiquarks and $N_{f}$ is the number of
effectively massless flavors. Evolution equations for the singlet
quark and  gluon distribution can be written as
\begin{eqnarray}
\frac{{\partial}G(x,Q^{2})}{{\partial}{\ln}Q^{2}}&=&{\int_{x}^{1}}dz[
P_{gg}(z,\alpha_{s}(Q^{2})) G(\frac{x}{z},Q^{2})\nonumber\\
&&+P_{gq}(z,\alpha_{s}(Q^{2})) F_{2}^{s}(\frac{x}{z},Q^{2})],\nonumber\\
\frac{{\partial}F_{2}^{s}(x,Q^{2})}{{\partial}{\ln}Q^{2}}&=&{\int_{x}^{1}}dz[
P_{qq}(z,\alpha_{s}(Q^{2})) F_{2}^{s}(\frac{x}{z},Q^{2})\nonumber\\
&&+2N_{f}P_{qg}(z,\alpha_{s}(Q^{2})) G(\frac{x}{z},Q^{2})].
\end{eqnarray}
where
$F_{2}^{s}(x,Q^{2})=(\sum_{i=1}^{N_{f}}e_{i}^{2}/N_{f})x[q_{i}(x,Q^{2})+\overline{q}_{i}(x,Q^{2})]$
and $G(x,Q^{2})=xg(x,Q^{2})$ are singlet and gluon distribution
functions. Some analytical solutions of the DGLAP evolution
equations have been reported in recent years [14-16] with
considerable phenomenological success.\\
In the evolution equations, the splitting functions $P_{ij}^{,}s$
are the LO, NLO and NNLO Altarelli- Parisi splitting kernels [17]
as
\begin{eqnarray}
P_{ij}(x,\alpha_{s}(Q^{2}))&=&\frac{\alpha_{s}(Q^{2})}{2\pi}P_{ij}^{\rm
LO}(x)+(\frac{\alpha_{s}(Q^{2})}{2\pi})^{2}P_{ij}^{\rm
NLO}(x)\nonumber\\
&&+(\frac{\alpha_{s}(Q^{2})}{2\pi})^{3} P_{ij}^{\rm NNLO}(x).
\end{eqnarray}
Here $P_{qq}, $$P_{qg}$, $P_{gq}$ and $P_{gg}$ are the
quark-quark, quark-gluon, gluon quark and gluon-gluon splitting
function respectively. Indeed $P_{qq}$  can be expressed as
$P_{qq}=P^{+}_{ns}+N_{f}(P_{qq}^{s}+P_{\overline{q}q}^{s}){\equiv}P^{+}_{ns}+P_{ps}$
which  the non-singlet splitting function $P^{+}_{ns}$ is
negligible at low-$x$ and can be ignored.  Therefore at low values
of $x$, the pure singlet term $P_{ps}$ dominates over
$P^{+}_{ns}$. Also the gluon-quark ($P_{gq}$) and quark-gluon
($P_{qg}$) are given by $P_{qg}=N_{f}P_{q_{i}g}$ and
$P_{gq}=P_{gq_{i}}$ where $P_{q_{i}g}$ and $P_{gq_{i}}$ are the
flavor-independent splitting
functions.\\
The running coupling constant ${\alpha_{s}}/{2\pi}$ at NNLO
analysis  has the following form as
\begin{eqnarray}
\frac{\alpha_{s}^{\rm
NNLO}}{2\pi}&=&\frac{2}{\beta_{0}t}[1-\frac{\beta_{1}{\ln}t}{\beta_{0}^{2}t}+\frac{1}{(\beta_{0}t)^{2}}
[(\frac{\beta_{1}}{\beta_{0}})^{2}\nonumber\\
&&(\ln^{2}t-{\ln}t+1)+\frac{\beta_{2}}{\beta_{0}}]],
\end{eqnarray}
where $\beta_{0}=\frac{1}{3}(33-2N_{f})$,
$\beta_{1}=102-\frac{38}{3}N_{f}$ and
$\beta_{2}=\frac{2857}{6}-\frac{6673}{18}N_{f}+\frac{325}{54}N_{f}^{2}$
are the one-loop,two-loop and three-loop corrections to the QCD
$\beta$-function. The variable $t$ is defined as
$t={\ln}(\frac{Q^{2}}{\Lambda^{2}})$ and $\Lambda$ is the QCD
cut- off parameter.\\
The power law behavior of singlet and gluon distribution functions
introduced as $F_{2}^{s}{\sim}x^{-\lambda_{s}}$ and
$G{\sim}x^{-\lambda_{g}}$.  The behavior of exponents, with a
$Q^{2}$ independent value, obeys the DGLAP equations when
$x^{-\lambda_{s,g}}\gg 1$. This behavior at small-$x$ is well
explained in terms of Regge-like ansatz [18]. In this region, the
Regge behavior of the singlet and gluon distributions are
corresponding to a pomeron exchange. Let us take the power law
behavior for distribution functions as
$F^{s}_{2}(x,Q^{2})=A_{s}(Q^{2})x^{-\lambda_{s}}$ and
$G(x,Q^{2})=A_{g}(Q^{2})x^{-\lambda_{g}}$. We note that exponents
$\lambda_{s}$ and $\lambda_{g}$ are given as the derivatives:
\begin{eqnarray}
\lambda_{s}=\frac{\partial \ln F_{2}^{s}(x,Q^{2})}{\partial
\ln(1/x)},\nonumber\\
 \lambda_{g}=\frac{\partial \ln
G(x,Q^{2})}{\partial \ln(1/x)}.
\end{eqnarray}
With respect to the DGLAP evolution equations (i.e. Eqs.8) and
used the Regge like behavior in r.h.s of Eqs.8 we have:
\begin{widetext}
\begin{eqnarray}
\frac{{\partial}G(x,Q^{2})}{{\partial}{\ln}Q^{2}}&=&{\int_{x}^{1}}dz[
P_{gg}(z,\alpha_{s}(Q^{2})) A_{g}(\frac{x}{z})^{-\lambda_{g}}
+P_{gq}(z,\alpha_{s}(Q^{2})) A_{s}(\frac{x}{z})^{-\lambda_{s}}],\nonumber\\
\frac{{\partial}F_{2}^{s}(x,Q^{2})}{{\partial}{\ln}Q^{2}}&=&{\int_{x}^{1}}dz[
P_{qq}(z,\alpha_{s}(Q^{2}))
A_{s}(\frac{x}{z})^{-\lambda_{s}}+2N_{f}P_{qg}(z,\alpha_{s}(Q^{2}))
A_{g}(\frac{x}{z})^{-\lambda_{g}}].
\end{eqnarray}
\end{widetext} These equations can be rearranged in the convolution
forms as we have
\begin{widetext}
\begin{eqnarray}
\frac{{\partial}G(x,Q^{2})}{{\partial}F_{2}^{s}(x,Q^{2})}=\frac{G(x,Q^{2})[P_{gg}(z,\alpha_{s}(Q^{2})){\otimes}
x^{\lambda_{g}}]+
F_{2}^{s}(x,Q^{2})[P_{gq}(z,\alpha_{s}(Q^{2})){\otimes}
x^{\lambda_{s}}]}{F_{2}^{s}(x,Q^{2})[P_{qq}(z,\alpha_{s}(Q^{2})){\otimes}
x^{\lambda_{s}}]+ G(x,Q^{2})[P_{qg}(z,\alpha_{s}(Q^{2})){\otimes}
x^{\lambda_{g}}]}.
\end{eqnarray}
\end{widetext}
Inserting Eqs.11 in l.h.s of Eq.13 then we obtain the ratio DGLAP
equations into an explicit relation between the singlet and gluon
distribution as
\begin{widetext}
\begin{eqnarray}
\frac{\lambda_{g}}{\lambda_{s}}\frac{G(x,Q^{2})}{F_{2}^{s}(x,Q^{2})}=\frac{G(x,Q^{2})[P_{gg}(x,\alpha_{s}(Q^{2})){\otimes}
x^{\lambda_{g}}]+
F_{2}^{s}(x,Q^{2})[P_{gq}(x,\alpha_{s}(Q^{2})){\otimes}
x^{\lambda_{s}}]}{F_{2}^{s}(x,Q^{2})[P_{qq}(x,\alpha_{s}(Q^{2})){\otimes}
x^{\lambda_{s}}]+ G(x,Q^{2})[P_{qg}(x,\alpha_{s}(Q^{2})){\otimes}
x^{\lambda_{g}}]},
\end{eqnarray}
\end{widetext}
where
\begin{eqnarray}
\frac{{\partial}G(x,Q^{2})}{{\partial}F_{2}^{s}(x,Q^{2})}&=&\frac{\lambda_{g}}{\lambda_{s}}\frac{G(x,Q^{2})}{F_{2}^{s}(x,Q^{2})}.
\end{eqnarray}
Here $\lambda_{s}$ and  $\lambda_{g}$ are taken as hard trajectory
intercepts minus one [19].\\
To solve  Eq.(14) one needs to define an relation between the
exponents and distribution functions as
$\lambda_{gs}=\lambda_{g}/\lambda_{s}$ and
$K(x,Q^{2})=G(x,Q^{2})/F_{2}^{s}(x,Q^{2})$  respectively [20-22].
Thus we can rewrite  Eq.(14) for obtain a general relation between
the singlet and gluon distribution functions. Therefore a
second-order
 equation is obtained for the ratio $K(x,Q^{2})$ in the following
 form
 \begin{eqnarray}
\lambda_{gs}D_{qg}(x,Q^{2})K^{2}(x,Q^{2})+[\lambda_{gs}C_{qq}(x,Q^{2})\nonumber\\
-A_{gg}(x,Q^{2})]K(x,Q^{2})-B_{gq}(x,Q^{2})=0,
 \end{eqnarray}
where, $A_{gg}(x,Q^{2})$, $B_{gq}(x,Q^{2})$, $C_{qq}(x,Q^{2})$ and
$D_{qg}(x,Q^{2})$ are given in Appendix A. Indeed Eq.16 leads to
the actual function form of
$K(x,Q^{2})$ for the ratio  $G/F_{2}^{s}$.\\
%%%%%%%%%%%%%%%%%%%%%%%%%%%%%%%%%%%%%%%%%%%%%%%%%%%%%%%%%%%%%%%%%%%%%%%%%%%%%
\subsection{2.2. The ratio $\frac{F_{L}(x,Q^{2})}{F_{2}(x,Q^{2})}$}
Now we consider the ratio $\frac{F_{L}(x,Q^{2})}{F_{2}(x,Q^{2})}$
with NNLO coefficient functions. In perturbative quantum
chromodynamics (pQCD), the longitudinal structure function is
proportional to hadronic tensor as it can be expressed by the
convolution of partonic structure functions. The longitudinal
structure function $F_{L}(x,Q^{2})$ of proton in terms of
coefficient functions can be written as [23]
\begin{eqnarray}
x^{-1}F_{L}=C_{L,ns}{\otimes}q_{ns}+<e^{2}>(C_{L,q}{\otimes}q_{s}+C_{L,g}{\otimes}g).\nonumber\\
\end{eqnarray}
The average squared charge is presented by $<e^{2}>$ and $q_{ns}$
stands for the usual flavor non-singlet contribution. This
contribution can be ignored safely at low-$x$ values and
$q_{s}=\sum_{N_{f}}(q+\overline{q})$ is the
flavor-singlet quark distribution.\\
The perturbative expansion of the coefficient functions can be
written as
\begin{eqnarray}
C_{L,q ~\&~
g}(\alpha_{s},x)=\sum_{n=1}(\frac{\alpha_{s}}{4\pi})^{n}c_{L,q
~\&~ g}(x).
\end{eqnarray}
Note that the coefficients up to NNLO are exhibited in compact
form in Ref.23 and also the singlet-quark coefficient function  is
decomposed into the nonsinglet and a pure singlet contribution.\\
On the basis of power-law behavior for the gluon and singlet
distribution functions, let us substitute this behavior in
Eq.(17). Thus Eq.(17) is reduced to the ratio
$\frac{F_{L}(x,Q^{2})}{F_{2}(x,Q^{2})}$ as we have
\begin{eqnarray}
\frac{F_{L}(x,Q^{2})}{F_{2}(x,Q^{2})}&=&[C_{L,q}(x,\alpha_{s}(Q^{2})){\otimes}
x^{\lambda_{s}}]\\
&&+<e^{2}>K(x,Q^{2})[C_{L,g}(x,\alpha_{s}(Q^{2})){\otimes}
x^{\lambda_{g}}],\nonumber
\end{eqnarray}
where $K(x,Q^{2})=G(x,Q^{2})/F_{2}^{s}(x,Q^{2})$ is taken from
Eq.(16). This equation (i.e., Eq.(19)) demonstrates the close
relation between the ratio structure functions and the color
dipole model
bound.\\
It is now possible to consider the proton structure function from
the reduced cross section on the right-hand side of Eq.(1).
Substituting Eq.(19) into Eq.(1), as
\begin{eqnarray}
F_{2}(x,Q^{2})=\sigma_{r}(x,Q^{2})[1-\frac{y^{2}}{Y_{+}}(Eq.19)]^{-1}.
\end{eqnarray}
This result shows that the proton structure function at $x-Q^{2}$
region can be determined using the kernels at high-order
corrections and the reduced cross section available data.\\
Typically, in HERA experimental data the ratio of the structure
functions is defined by $R(x,Q^{2})$, as
${F_{L}}/{F_{2}}={R}/{(1+R)}$. Also  it may be achieved this ratio
via the DGLAP combined evolution equations. Therefore the proton
structure function is expected to determine (using Eqs.19-20) at
some of points which did not report in experimental data. This
procedure requires only the reduced cross section and kernels
in evolution equations with respect to the effective intercepts.\\
%%%%%%%%%%%%%%%%%%%%%%%%%%%%%%%%%%%%%%%%%%%%%%%%%%%%%%%
\subsection{3. Result and Discussion}
In this paper, we obtain the ratio $G(x,Q^{2})/F_{2}^{s}(x,Q^{2})$
and $F_{L}(x,Q^{2})/F_{2}(x,Q^{2})$ and the proton structure
function at NNLO analysis respectively. The analysis is performed
in the range $10^{-5}\leq x \leq 10^{-2}$ and $1.5\leq Q^{2} \leq
150~GeV^{2}$. We should first extract the ratio $G/F_{2}^{s}$ in
Fig.1 with $\tau$ variable where
$\tau=\frac{Q^{2}}{Q_{0}^{2}}(\frac{x}{x_{0}})^{-\lambda}$. Here
$Q_{0}^{2}=1~GeV^{2}$, $x_{0}=3.0\times 10^{-4}$ and
$\lambda=\lambda_{s}$. The effective exponents for gluon and
singlet distributions are defined with an exponent of
$\lambda_{g}=0.424$ and $\lambda_{s}=0.327$ respectively [18].
These values are compatible with other results [19]. Also this
ratio is plotted in Fig.2 with $Q^{2}$ at a certain representation
value of fixed $x$.\\
In what follows the ratio $F_{L}(x,Q^{2})/F_{2}(x,Q^{2})$, with
respect to Eqs.(16) and (19), is calculated and presented in
Fig.3. In this figure the ratio of the structure functions
compared with the H1 data [1] and with the result obtained by the
color dipole model bound [10]. The error bars of the ratio
$\frac{F_{L}}{F_{2}}$ are determined by the following form [11]
\begin{eqnarray}
\Delta(\frac{F_{L}}{F_{2}})=\frac{F_{L}}{F_{2}}\sqrt{(\frac{\Delta
F_{L} }{F_{L}})^{2} +(\frac{\Delta F_{2} }{F_{2}})^{2}},
\end{eqnarray}
where $\Delta F_{L}$ and  $\Delta F_{2}$ are collected from the H1
experimental data in Ref.[1]. The good agreement  between this
method and the experimental data indicates that our results has a
bound asymptotic behavior and it is compatible with the color
dipole model bound.\\
In Ref.[3] the measured structure functions $F_{L}(x,Q^{2})$ for
$Q^{2}\leq 35~ GeV^{2}$ with total uncertainties below $0.3$ and
$Q^{2}=45~ GeV^{2}$ below $0.4$ are presented. The ratio
$R(=\frac{F_{L}}{F_{2}-F_{L}})$ is found at $R=0.260{\pm}0.050$
which this value is constant at the region $7.10^{-5}<x<2.10^{-3}$
and $3.5\leq Q^{2} \leq 45 GeV^{2}$. We know that the color dipole
model has been described for virtual photon-proton scattering at
low $x$ and low $Q^{2}$ values. In color dipole model the ratio
$R$ lead to the bound $R\leq 0.372$ [7]. This value decrease when
another approaches were developed to describe the dipole-proton
cross section, such as IIM and B-SAT. Where the first one is a
model based on the colour glass condensate approach to the high
parton density regime and the another one is a model with the
generalised impact parameter dipole saturation respectively [9].
In Ref.[24] ZEUS Collaboration is shown that the overall value of
$R$ from both the unconstrained and constrained fits is $R =
0.105^{+0.055} _{-0.037}$ in wide range of $Q^{2}$ values ($5\leq
Q^{2} \leq 110~GeV^{2}$). In both the DGLAP and the dipole models
the structure function $F_{L}$ and the ratio of the structure
functions  can be calculated. It is thus of interest to compare
predictions of the different models with the data.  For high
$Q^{2} > 10~ GeV^{2}$, theses models agree with the data but for
lower $Q^{2}$ values, there is a
significant difference between the predictions.\\
Indeed authors of Ref.[11] have shown that for realistic
dipole-proton cross-section the bound is reduced from $0.27$ to
$0.22$. With respect to this method we can see from Fig.3 which
the ratio structure functions  lie below the EMNS bound at
moderate and high values of $\tau$,  and it is comparable with the
EMNS bound at low values of $\tau$. In this region the data
indicate a decrease of the ratio $\frac{F_{L}}{F_{2}}$ for small
values of $Q^{2}$, as we do not expect for evolution equation
predicted with respect to the singlet and gluon distribution
behavior. This behavior allows us to speculate that there may be a
need for QCD resummations beyond the conventional DGLAP equations
or the need for non-linear evolution equations which take account
of gluon recombination and the possibility of gluon saturation.
Such effects can be described by non-linear evolution equations
including higher-twist corrections at low -$x$ values [25-26]. The
expectation is that such terms are important for the longitudinal
structure function but not for the structure function $F_{2}$.\\
Indeed  the introduction of higher-twist terms is one possible way
to extend the DGLAP framework to low $Q^{2}$ values. Such terms
have been introduced at low-$x$ values since, for the kinematics
of HERA, low $Q^{2}$ is only accessed at low $x$. To better
illustrate our calculations at low $Q^{2}$, we added a higher
twist term in the description of the structure functions, for
 HERA data on deep inelastic scattering, at low $x$
and low $Q^{2}$ values. It can be clearly seen that our
predictions with respect to the higher twist (HT) analyses are
comparable with data at this region. The leading twist
perturbative QCD predictions of the structure functions $F_{2}$
and $F_{L}$ augment by a simple higher twist term such that
\begin{eqnarray}
F_{2}^{HT}=F_{2}^{DGLAP}(1+\frac{A_{2}}{Q^{2}}),\nonumber\\
F_{L}^{HT}=F_{L}^{DGLAP}(1+\frac{A_{L}}{Q^{2}}),
\end{eqnarray}
where $A^{HT}_{2}=0.12~\pm~0.07~GeV^{2}$ and
$A^{HT}_{L}=5.5~\pm~0.6~GeV^{2}$ are free parameters at NNLO
[25-26]. Using the HT terms in Eqs.16 and 19, we can evaluate the
HT corrections to the ratio $G/F_{2}^{s}$ and $F_{L}/F_{2}$ as we
have
\begin{eqnarray}
\mathrm{Eq.(16)}&\Rightarrow &\lambda_{gs}D_{qg}(x,Q^{2})K^{2}(x,Q^{2})\frac{1}{1+A^{HT}_{2}/Q^{2}}\nonumber\\
&&+[\lambda_{gs}C_{qq}(x,Q^{2})-A_{gg}(x,Q^{2})]K(x,Q^{2})\nonumber\\
&&-(1+A^{HT}_{2}/Q^{2})B_{gq}(x,Q^{2})=0,
\end{eqnarray}
and
\begin{eqnarray}
\mathrm{Eq.(19)}&\Rightarrow &
\frac{F_{L}(x,Q^{2})}{F_{2}(x,Q^{2})}=\frac{1}{1+A^{HT}_{L}/Q^{2}}\{(1+A^{HT}_{2}/Q^{2})\nonumber\\
&&[C_{L,q}(x,\alpha_{s}(Q^{2})) {\otimes} x^{\lambda_{s}}]
+<e^{2}>K(x,Q^{2})\nonumber\\
&&[C_{L,g}(x,\alpha_{s}(Q^{2})){\otimes} x^{\lambda_{g}}]\}.
\end{eqnarray}
The results are shown in Fig.4. In this figure the HT predictions
of the ratio $\frac{F_{L}}{F_{2}}$ presented on the H1
measurement, where a decrease of the ratio for small $Q^{2}$
values is observable. Comparison of the DGLAP data with HT data
are shown in Fig.5.\\
Having checked that the ratio method obtained reproduced
satisfactorily the existing DIS reduced cross section on the
electron-proton collisions in upper $x$ domain at fixed $Q^{2}$
values. We shall now use it to make predictions for these values
as expected there are no data for electron- proton collisions in
this region. The results are depicted in Table I. In this table we
compared results with Ref.4 data and obtained the starting
$x-Q^{2}$ points as this method can be prediction data at
extrapolation points . These results are comparable with
literature as accompanied with total errors. In figure 6, the
proton structure function have been shown and compared at
$Q^{2}=20~GeV^{2}$ with H1 2001 and H1 2014 data [1,4]
respectively. These figures indicates that the obtained results
from present analysis based on DGLAP bound are in good agreements
with the ones obtained by HERA data. A comparison has also been
shown at $Q^{2}=8.5~GeV^{2}$ in figure 7 and compared with H1 2011
data [3] as accompanied with total errors. We see that all
predictions are consistent with the large $y$ data.\\
%%%%%%%%%%%%%%%%%%%%%%%%%%%%%%%%%%%%%%%%%%%%%%%%%%%%%%%%%%%%
\subsection{4. Heavy flavor contribution}
As our further research activities we hope to study the ratio of
structure functions to get an analytical solutions for heavy quark
contributions of the structure functions. When the virtual photon
interacts indirectly with a gluon in the proton then a heavy quark
pair produced via the direct boson-gluon fusion processes. At
low-$x$ this behavior is related to the growth of gluon
distribution via the $g \rightarrow q_{H}\overline{q}_{H}~(H=c,b)$
transition [27-28]. Then the perturbative predictions for
$F_{L}(x,Q^{2})$ at the $N_{f} = 3$ light quark flavor sector and
heavy contribution can be written as
\begin{eqnarray}
x^{-1}F_{L}=C_{L,ns}{\otimes}q_{ns}+\frac{2}{9}(C_{L,q}{\otimes}q_{s}+C_{L,g}{\otimes}g)+ x^{-1}F_{L}^{H}.\nonumber\\
\end{eqnarray}
Also the heavy quark contribution to the total structure functions
is  $F_{2}(x,Q^{2})=F_{2}^{\mathrm{light}}
+F_{2}^{\mathrm{Heavy}}$ where `light' refers to the common $u; d;
s$ (anti)quarks and gluon initiated contributions at fixed flavor
number scheme. The heavy quark contributions at small-$x$ are
given by
\begin{eqnarray}
F_{k}^{H}(x,Q^{2})&=&\frac{Q^{2}\alpha_{s}(\mu^{2})}{4\pi^{2}m^{2}_{H}}\int_{ax}^{1}\frac{dy}{y}yg(y,\mu^{2})e^{2}_{H}\{C_{k,g}^{(0)}
\nonumber\\
&&+4\pi\alpha_{s}(\mu^{2})[C_{k,g}^{(1)}+\overline{C}_{k,g}^{(1)}\ln\frac{\mu^{2}}{m^{2}_{H}}]+...\}\nonumber\\
&&,~(k=2,L).
\end{eqnarray}
Here $e_{H}$ denotes the heavy charge and $m_{H}$ denotes the
heavy quark mass [2]. The lower limit of integration is given by
$y_{min}=ax=(1+4\frac{m_{H}^{2}}{Q^{2}})x$ and the mass
factorization scale $\mu$ which has been put equal to the
renormalization scale is assumed to be either $\mu^{2} =
4m^{2}_{H}$ or $\mu^{2}= 4m^{2}_{H}+Q^{2}$.\\
Now discuss the bound for the ratio $F_{L}/F_{2}$ for
$n_{f}=3+\mathrm{Heavy}$, as
\begin{eqnarray}
\frac{F_{L}^{\mathrm{light}+c}}{F_{2}^{\mathrm{light}+c}}&=&\frac{F_{L}^{N_{f}=3}+F_{L}^{c}}{F_{2}^{N_{f}=3}+F_{2}^{c}}\nonumber\\
&&=\frac{F_{L}^{N_{f}=3}/F_{2}^{N_{f}=3}+F_{L}^{c}/F_{2}^{N_{f}=3}}{1+F_{2}^{c}/F_{2}^{N_{f}=3}},
\end{eqnarray}
when $m_{c}^{2}<\mu^{2}<m_{b}^{2}$ and
\begin{eqnarray}
\frac{F_{L}^{\mathrm{light}+c+b}}{F_{2}^{\mathrm{light}+c+b}}&=&\frac{F_{L}^{N_{f}=3}+F_{L}^{c}+F_{L}^{b}}{F_{2}^{N_{f}=3}+F_{2}^{c}+F_{2}^{b}}\\
&&=\frac{F_{L}^{N_{f}=3}/F_{2}^{N_{f}=3}+F_{L}^{c}/F_{2}^{N_{f}=3}+F_{L}^{b}/F_{2}^{N_{f}=3}}{1+F_{2}^{c}/F_{2}^{N_{f}=3}+F_{2}^{b}/F_{2}^{N_{f}=3}},\nonumber
\end{eqnarray}
when $m_{b}^{2}<\mu^{2}<m_{t}^{2}$. The ratio
$\frac{F_{L}^{N_{f}}}{F_{2}^{N_{f}}}$ not only  mentioned earlier
in Eq.(19) for $N_{f}=4$ but also it will be consider here for
$N_{f}=3$.\\
The ratio of the heavy flavor structure functions are described by
a convolution between the gluon distribution and the heavy Wilson
coefficients as we have
\begin{eqnarray}
\frac{F_{k}^{H}}{F_{2}^{N_{f}=3}}&=&\frac{G(x,\mu^{2})}{F_{2}^{N_{f}=3}}\{C_{k,g}^{H}(x,\frac{Q^{2}}{\mu^{2}})\otimes
x^{\lambda_{g}}\}\nonumber\\
&&=K(x,Q^{2})[\mathrm{Eq.16 ~for~
N_{f}=3}]\{C_{k,g}^{H}(x,\frac{Q^{2}}{\mu^{2}})\otimes
x^{\lambda_{g}}\}.\nonumber\\
\end{eqnarray}
In Table II the effects of heavy quarks on the ratio of structure
functions are considered. We observe that the bottom quark effect
on the ratio is negligible in the wide range of $Q^{2}$ values. In
the present analysis we use $m_{c}=1.3~GeV$ and $m_{b}=4.5~GeV$,
therefore
\begin{eqnarray}
\frac{F_{L}^{\mathrm{light}+c+b}}{F_{2}^{\mathrm{light}+c+b}}{\simeq}\frac{F_{L}^{\mathrm{light}+c}}{F_{2}^{\mathrm{light}+c}}.
\end{eqnarray}
In Table III we observe that the bound is changed when the
inclusion charm mass effects is going in the bound. The charm
effects in the bound of ratio show that in the wide range of
$Q^{2}$ values we have this behavior as
\begin{eqnarray}
\frac{F_{L}}{F_{2}}|_{{N_{f}=3+c}} ~
{\leq}~\frac{F_{L}}{F_{2}}|_{{N_{f}=4}}.
\end{eqnarray}
We note that the ratio $F_{L}/F_{2}$ in $N_{f}=3+c$ is
approximately equivalent to the ratio in $N_{f}=4$ at high-$Q^{2}$
values. A comparison of the various contributions to the ratio
shows that for low-$Q^{2}$ values the charm quark contribution in
the ratio is about $12\%$ or less. Therefore the charm effect in
the bound decrease as $Q^{2}$ increases. Indeed the ratio bound is
lower than EMNS bound when the charm mass effects are taken into
account.\\

\subsection{5. Conclusion}
In this paper we have found that there is in general an analytical
relation between the gluon distribution function and singlet
structure function at low $x$ region into the effective exponents.
The ratio of the structure functions into the DGLAP evolution
equations at small $x$ at NNLO analysis is studied and compared
with EMNS bound in this region. Results are comparable with the
experimental data and they are lower than EMNS bound at
high-$Q^{2}$ values. Our results are very close to the bounds for
low-$Q^{2}$ values as we have discussed the meaning of these
findings from the points of view of higher twist terms added to
the structure functions. Having checked that this model gives a
good description of the ratio $F_{L}/F_{2}$ then we predict
$F_{2}(x,Q^{2})$ with respect to the reduced cross section
measured in HERA collisions. We observed that the general
solutions are comparable with the available experimental data.
Finally we discussed the
charm quark effects in bounds at high and low-$Q^{2}$ values.\\
%%%%%%%%%%%%%%%%%%%%%%%%%%%%%%%%%%%%%%%%%%%%%%%%%%%%%%%%%%
\subsection{Appendix A}
The explicit forms of the functions $A_{gg}(x,Q^{2})$,
$B_{gq}(x,Q^{2})$, $C_{qq}(x,Q^{2})$ and $D_{qg}(x,Q^{2})$ are
defined by
\begin{eqnarray}
A_{gg}(x,Q^{2})&=&P_{gg}(x,\alpha_{s}(Q^{2})){\otimes}
x^{\lambda_{g}}\nonumber\\
 &&{\equiv}{\int_{x}^{1}}dz
P_{gg}(z,\alpha_{s}(Q^{2})) {z}^{\lambda_{g}},\nonumber\\
B_{gq}(x,Q^{2})&=&P_{gq}(x,\alpha_{s}(Q^{2})){\otimes} x^{\lambda_{s}}\nonumber\\
&&{\equiv}{\int_{x}^{1}}dz P_{gq}(z,\alpha_{s}(Q^{2})) {z}^{\lambda_{s}},\nonumber\\
C_{qq}(x,Q^{2})&=&P_{qq}(x,\alpha_{s}(Q^{2})){\otimes}
x^{\lambda_{s}}\nonumber\\
&&{\equiv}{\int_{x}^{1}}dz
P_{qq}(z,\alpha_{s}(Q^{2})){z}^{\lambda_{s}},\nonumber\\
D_{qg}(x,Q^{2})&=&P_{qg}(x,\alpha_{s}(Q^{2})){\otimes}
x^{\lambda_{g}}\nonumber\\
&&{\equiv}{\int_{x}^{1}}dz
2N_{f}P_{qg}(z,\alpha_{s}(Q^{2})){z}^{\lambda_{g}}.
\end{eqnarray}
where the strong coupling constant $\alpha_{s}$ and splitting
functions up to NNLO are given in Ref.[14].
%%%%%%%%%%%%%%%%%%%%%%%%%%%%%%%%%%%%%%%%%%%%%%%%%%%%%%%%%%%%%
\subsection{References}

1. V. Andreev  et al. [H1 Collab.], Eur. Phys. J. C{\bf74}(2014)2814.\\
2. H. Abramowicz et al. [H1 and ZEUS Collab.], Eur. Phys. J. C {\bf75}(2015)580.\\
3. F.D. Aaron et al. [H1 Collaboration], phys.Lett.B\textbf{665},
139(2008); Eur.Phys.J.C\textbf{71},1579(2011).\\
4. C.Adloff et al. [H1 Collaboration], Eur.Phys.J.C\textbf{21}, 33(2001).\\
5. V.Tvaskis et al., Phys.Rev.C\textbf{97}, 045204(2018).\\
6. N.N.Nikolaev and B.G.Zakharov, Z.Phys.C\textbf{49}, 607(1991);
Z.Phys.C\textbf{53}, 331(1992).\\
7. C.Ewerz and O.Nachtmann, Phys.Lett.B\textbf{648}, 279(2007).\\
8. C.Ewerz, A. von Manteuffel and O.Nachtmann, Phys.Rev.D\textbf{77}, 074022(2008).\\
9. K.Golec-Biernat and W$\mathrm{\ddot{u}}$sthoff, Phys.Rev.D\textbf{59}, 014017(1999);
E. Iancu, K. Itakura, and S. Munier, Phys. Lett. B\textbf{590}, 199(2004); H. Kowalski, L.Motyka, and G.Watt, Phys. Rev. D\textbf{74}, 074016(2006).\\
10. C.Ewerz et al., Phys.lett.B\textbf{720}, 181(2013).\\
11. M.Niedziela and M.Praszalowicz, Acta Phys.Polon. B{\bf46}, 2019(2015).\\
12. Yu.L.Dokshitzer, Sov.Phys.JETP {\textbf{46}}, 641(1977);
G.Altarelli and G.Parisi, Nucl.Phys.B \textbf{126}, 298(1977);
V.N.Gribov and L.N.Lipatov, Sov.J.Nucl.Phys. \textbf{15},
438(1972).\\
13. G.Altarelli and G.Martinelli, Phys.Lett. B{\bf76}, 89(1978).\\
14. G. R. Boroun and B. Rezaei, Eur. Phys. J. C{\bf73}, 2412(2013); G. R. Boroun and B. Rezaei, Eur. Phys. J. C{\bf72}(2012)2221.\\
15. S.Shoeibi, et al., Phys.Rev.D{\bf7}, 074013 (2018); H.Khanpour
et al., Eur.Phys. J.C{\bf78}, 7(2018); S.M. Mossavi Nejad et al.,
Phys.Rev.C{\bf94}, 045201(2016);
 F.Taghavi-Shahri et al., Phys.Rev.D{\bf93}, 114024 (2016); H.Khanpour et al.,
  Phys.Rev.C{\bf95}, 035201(2017); J.Sheibaniet al.,
  Phys.Rev.C{\bf98}, 045211(2018).\\
16. G.R.Boroun Phys.Rev.C{\bf97}, 015206 (2018); B.Rezaei and
G.R.Boroun, arXiv:1811.02785(2018).\\
17. A. Vogt et al., Nucl. Phys. B{\bf691}(2004)129.\\
18. P.D.B.Collins, Cambridge University Press, Cambridge (1977).\\
19. N. N. Nikolaev and W. Schäfer, Phys. Rev. D {\bf74}, 014023
(2006); E.Gotsman et al., arXiv:1712.06992(2017); H.Kowalski et al., arXiv:1707.014460(2017).\\
20. G. R. Boroun, Eur. Phys. J. A{\bf50}, 69(2014).\\
21. M. Devee and J. K. Sarma, Eur. Phys. J. C{\bf72}, 2036(2012).\\
22. L. Machahari and D.K.Choudhury,
Eur.Phys.J.A{\textbf{54}}, 69(2018).\\
23. S. Moch et al., Phys. Lett. B{\bf606}, 123(2005).\\
24. H.Abromowicz et al. [ZEUS Collaboration],
Phys.Rev.D\textbf{9}, 072002(2014).\\
25. A.M.Cooper-Sarkar, arXiv:1605.08577v1 [hep-ph] 27 May 2016; I.Abt et.al., arXiv:1604.02299v2 [hep-ph] 11 Oct 2016.\\
26.F.D. Aaron et al. [H1 Collaboration], Eur.Phys.J. C{\bf63}, 625(2009).\\
27. N.N.Nikolaev and V.R.Zoller, Phys.Atom.Nucl\textbf{73},
672(2010); A. Y. Illarionov and A. V. Kotikov, Phys.Atom.Nucl.
{\bf75}, 1234 (2012); N.Ya.Ivanov, and B.A.Kniehl, Eur.Phys.J.C\textbf{59}, 647(2009); L.P.Kaptari et al., arXiv:1812.00361[hep-ph](2018).\\
28. G.R.Boroun, B.Rezaei, JETP,Vol.115, No.7, PP.427 (2012);
Nucl.Phys.B{\bf857}, 143(2012); Eur.Phys.J.C{\bf72}, 2221 (2012);
EPL{\bf100},41001(2012); Nucl.Phys.A{\bf929}, 119(2014); G.R.Boroun, Nucl.Phys.B{\bf884}, 684(2014).\\

%%%%%%%%%%%%%%%%%%%%%%%%%%%%%%%%%%%%%%%%%%%%%%%%%%%%%%%%%%
\begin{table}
\centering \caption{The proton structure function determined based
on the reduced  cross section data that accompanied with total
errors. }\label{table:table1}
\begin{minipage}{\linewidth}
\renewcommand{\thefootnote}{\thempfootnote}
\centering
\begin{tabular}{|l||c|c||c|c||c|c|} \hline\noalign{\smallskip} $Q^{2}(GeV^{2})$ & $ x$ &
$ y$ & $ \widetilde{\sigma}$ & $ \Delta(\%)$
& $F_{2}$(Ref.4) &$F_{2}$   \\
\hline\noalign{\smallskip}
2  & 0.0000327 & 0.675 & 0.805 & 7.4 & ---- & 0.911    \\
2  & 0.0000500 & 0.442 & 0.823 & 3.5 & 0.851 & 0.859    \\
2.5& 0.0000409 & 0.675 & 0.899 & 7.4 & ---- & 1.009    \\
2.5& 0.0000500 & 0.552 & 0.859 & 3.7 & 0.909 & 0.921    \\
3.5& 0.0000573 & 0.675 & 0.897 & 7.0 & ---- & 0.955    \\
3.5& 0.0000800 & 0.483 & 0.925 & 2.9 & 0.964 & 0.971    \\
5& 0.0000818 & 0.675 & 1.019 & 6.6 & ---- & 1.118    \\
5& 0.000130 & 0.425 & 1.015 & 2.4 & 1.043 & 1.045    \\
8.5& 0.000139 & 0.675 & 1.097 & 4.9 & ---- & 1.186    \\
8.5& 0.000200 & 0.470 & 1.152 & 2.9 & 1.193 & 1.189    \\
12& 0.000161 & 0.825 & 1.226 & 5.8 & ---- & 1.377    \\
12& 0.000197 & 0.675 & 1.269 & 3.5 & ---- & 1.362    \\
12& 0.000320 & 0.415 & 1.217 & 2.0 & 1.249 & 1.243    \\
15& 0.000201 & 0.825 & 1.255 & 5.2 & ---- & 1.399    \\
15& 0.000246 & 0.675 & 1.361 & 3.3 & ---- & 1.454    \\
15& 0.000320 & 0.519 & 1.283 & 2.4 & 1.342 & 1.328    \\
20& 0.000268 & 0.825 & 1.313 & 5.2 & ---- & 1.451    \\
20& 0.000328 & 0.675 & 1.383 & 2.7 & ---- & 1.470    \\
20& 0.000500 & 0.443 & 1.285 & 2.0 & 1.324 & 1.313    \\
25& 0.000335 & 0.825 & 1.379 & 5.9 & ---- & 1.515    \\
25& 0.000410 & 0.675 & 1.371 & 2.6 & ---- & 1.452    \\
25& 0.000500 & 0.553 & 1.345 & 2.4 & 1.417 & 1.393    \\
35& 0.000574 & 0.675 & 1.473 & 2.7 & ---- & 1.553    \\
35& 0.000800 & 0.484 & 1.354 & 2.2 & 1.405 & 1.386    \\
\hline\noalign{\smallskip}
\end{tabular}
\end{minipage}
\end{table}
%%%%%%%%%%%%%%%%%%%%%%%%%%%%%%%%%%%%%%%%%%%%%%%%%%%%%%%%%%

%%%%%%%%%%%%%%%%%%%%%%%%%%%%%%%%%%%%%%%%%%%%%%%%%%%%%%%%%%
\begin{figure}
\centering
\includegraphics[width=0.5\textwidth]{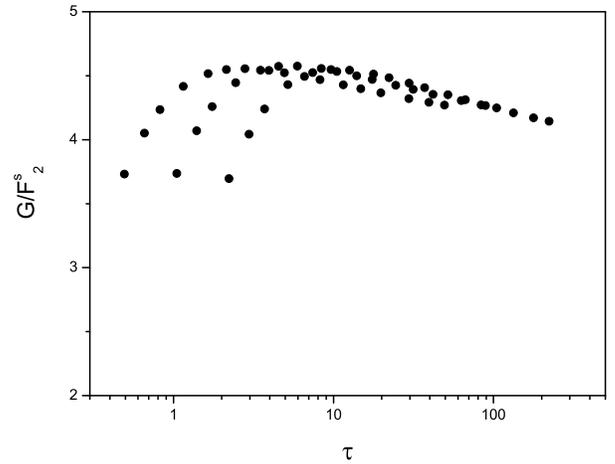}
\caption{Plot of the ratio $G/F_{2}^{s}$ vis $\tau$.} \label{Fig1}
\end{figure}
\begin{figure}
\centering
\includegraphics[width=0.5\textwidth]{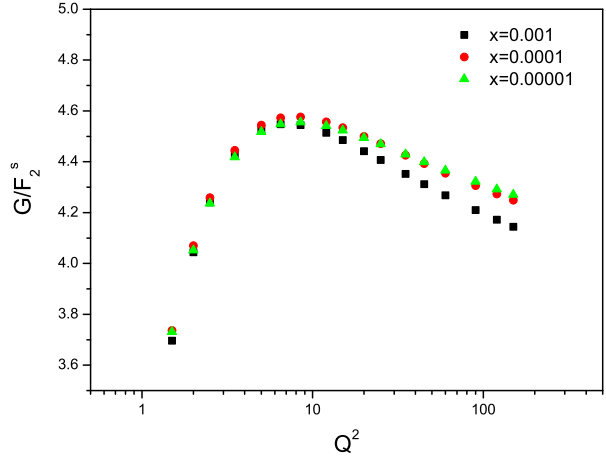}
\caption{Plot of the ratio $G/F_{2}^{s}$ as a function of
$Q^{2}$.} \label{Fig2}
\end{figure}
\begin{figure}
\centering
\includegraphics[width=0.5\textwidth]{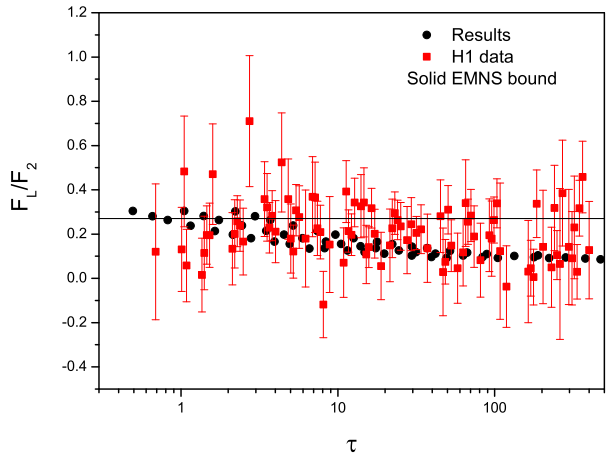}
\caption{Ratio $F_{L}/F_{2}$ plotted as function of scaling
variable $\tau$ compared with H1 data [1]. Straight line
corresponds to the color dipole model bound [10].} \label{Fig3}
\end{figure}
\begin{figure}
\centering
\includegraphics[width=0.5\textwidth]{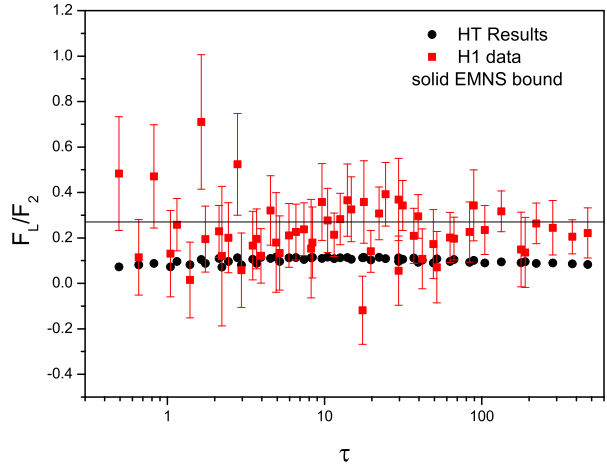}
\caption{The HT data of the ratio plotted as function of scaling
variable $\tau$ compared with H1 data [1]. Straight line
corresponds to the color dipole model bound [10].} \label{Fig3}
\end{figure}
\begin{figure}
\centering
\includegraphics[width=0.5\textwidth]{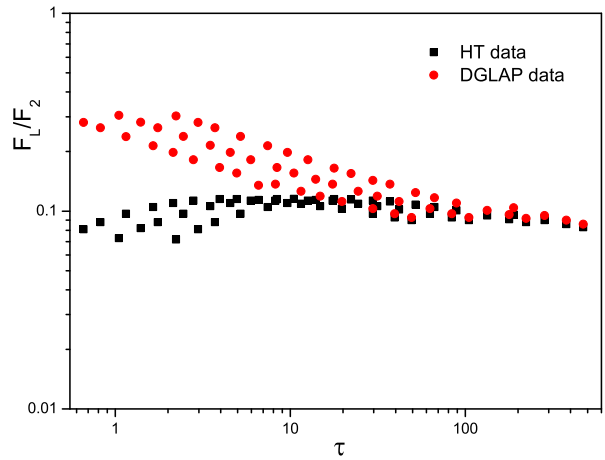}
\caption{The DGLAP and HT data for the ratio plotted as function
of $\tau$.} \label{Fig3}
\end{figure}
\begin{figure}
\centering
\includegraphics[width=0.5\textwidth]{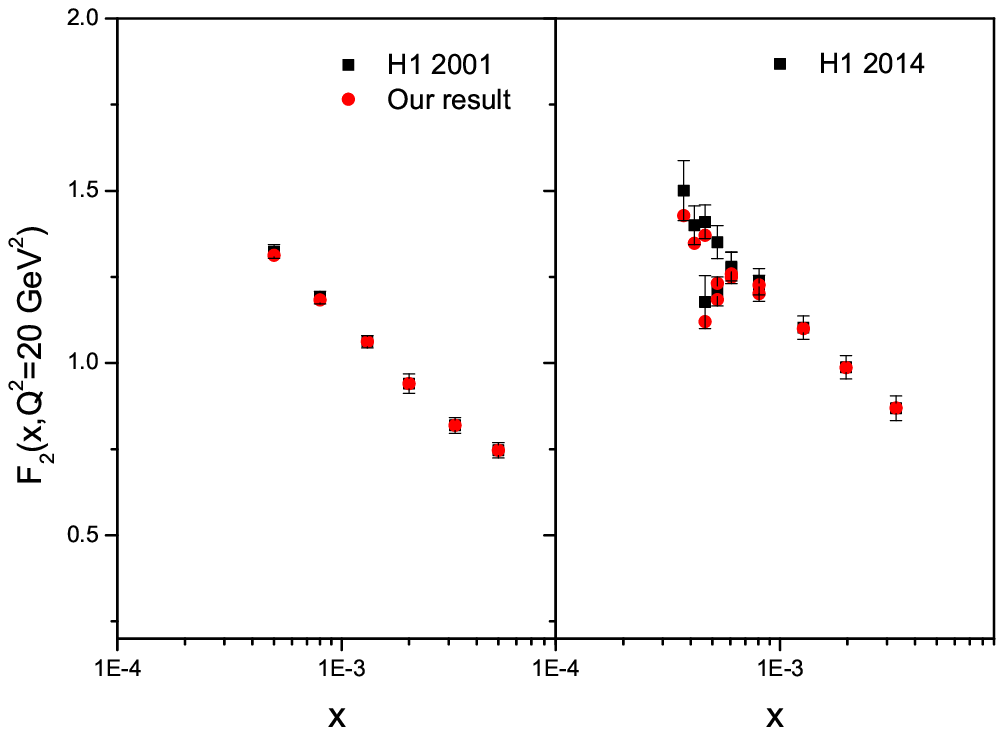}
\caption{The proton structure function $F_{2}(x,Q^{2})$ for
$Q^{2}=20 ~GeV^{2}$ compared  with H1 data [1, 4] where
accompanied with total errors.} \label{Fig3}
\end{figure}
\begin{figure}
\centering
\includegraphics[width=0.5\textwidth]{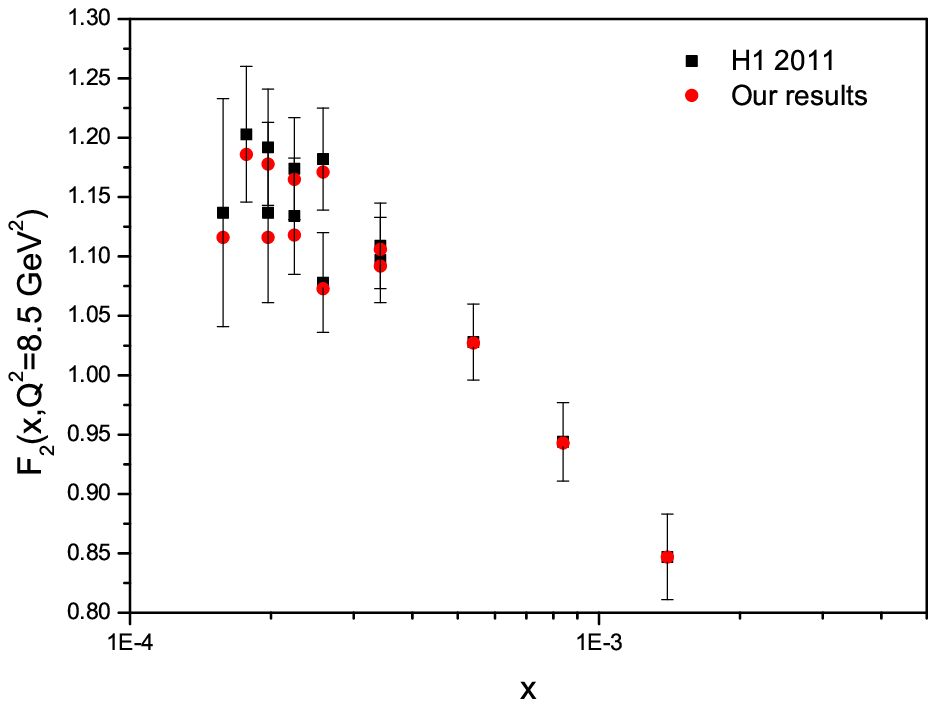}
\caption{The proton structure function $F_{2}(x,Q^{2})$ for
$Q^{2}=8.5~ GeV^{2}$ compared  with H1 data [3] where accompanied
with total errors.} \label{Fig3}
\end{figure}
%%%%%%%%%%%%%%%%%%%%%%%%%%%%%%
%%%%%%%%%%%%%%%%%%%%%%%%%%%%%%%%%%%%%%%%%%%%%%%%%%%%%%%%%%
\begin{table}
\centering \caption{The ratio $F_{k}^{H}/F_{2}$ determined in the
case $N_{f}=3$  for different values of $Q^{2}$.
}\label{table:table2}
\begin{minipage}{\linewidth}
\renewcommand{\thefootnote}{\thempfootnote}
\centering
\begin{tabular}{|l||c|c||c|c|} \hline\noalign{\smallskip} $Q^{2}(GeV^{2})$ & $F_{2}^{b}/F_{2}$ &
$ F_{L}^{b}/F_{2}$ & $F_{2}^{c}/F_{2}$ & $F_{L}^{c}/F_{2}$
  \\
\hline\noalign{\smallskip}
2   & 0.00008 & 0.0000005 & 0.013 & 0.0008  \\
20  & 0.0016 & 0.00009 & 0.079 & 0.015    \\
200 & 0.012 & 0.002 & 0.130 & 0.025    \\
400 & 0.015 & 0.003 & 0.138 & 0.025    \\
\hline\noalign{\smallskip}
\end{tabular}
\end{minipage}
\end{table}
%%%%%%%%%%%%%%%%%%%%%%%%%%%%%%%%%%%%%%%%%%%%%%%%%%%%%%%%%%
%%%%%%%%%%%%%%%%%%%%%%%%%%%%%%%%%%%%%%%%%%%%%%%%%%%%%%%%%
\begin{table}
\centering \caption{The ratio $F_{L}/F_{2}$ determined in the case
$N_{f}=3$ and $N_{f}=3+\mathrm{charm}$ and compared with the case
$N_{f}=4$ for different values of $Q^{2}$. }\label{table:table3}
\begin{minipage}{\linewidth}
\renewcommand{\thefootnote}{\thempfootnote}
\centering
\begin{tabular}{|l||c|c||c|} \hline\noalign{\smallskip} $Q^{2}(GeV^{2})$ &
 $F_{L}/F_{2}|_{N_{f}=3}$ & $F_{L}/F_{2}|_{N_{f}=3+c}$ & $F_{L}/F_{2}|_{N_{f}=4}$
  \\
\hline\noalign{\smallskip}
1.5   & 0.275 & 0.273 & 0.305 \\
2  & 0.247 & 0.245  & 0.281  \\
4 & 0.192 & 0.189    & 0.229 \\
10 & 0.142 & 0.143    & 0.174 \\
20 & 0.116 & 0.122     & 0.144\\
50 & 0.093 & 0.104     & 0.116\\
90 & 0.083 & 0.096    & 0.103 \\
120 & 0.078 & 0.092    & 0.097 \\
150 & 0.075 & 0.090    & 0.093 \\
200 & 0.071 & 0.086    & 0.088 \\
 \hline\noalign{\smallskip}
\end{tabular}
\end{minipage}
\end{table}
%%%%%%%%%%%%%%%%%%%%%%%%%%%%%%%%%%%%%%%%%%%%%%%%%%%%%%%%%%
\end{document}